\documentstyle[12pt,psfig,draft,rotate]{nature}
\def\simlt{\mathrel{\hbox{\rlap{\hbox{\lower4pt\hbox{$\sim$}}}\hbox{$<$}}}}
\def\simgt{\mathrel{\hbox{\rlap{\hbox{\lower4pt\hbox{$\sim$}}}\hbox{$>$}}}}
\makeatletter
\def\@xfloat#1[#2]{\ifhmode \@bsphack\@floatpenalty -\@Mii\else
   \@floatpenalty-\@Miii\fi\def\@captype{#1}\ifinner
      \@parmoderr\@floatpenalty\z@
    \else\@next\@currbox\@freelist{\@tempcnta\csname ftype@#1\endcsname
       \multiply\@tempcnta\@xxxii\advance\@tempcnta\sixt@@n
       \@tfor \@tempa :=#2\do
                        {\if\@tempa h\advance\@tempcnta \@ne\fi
                         \if\@tempa t\advance\@tempcnta \tw@\fi
                         \if\@tempa b\advance\@tempcnta 4\relax\fi
                         \if\@tempa p\advance\@tempcnta 8\relax\fi
         }\global\count\@currbox\@tempcnta}\@fltovf\fi
    \global\setbox\@currbox\vbox\bgroup 
    \def\baselinestretch{1.5}\@normalsize
    \boxmaxdepth\z@
    \hsize\columnwidth \@parboxrestore}
\makeatother
\begin{document}
\title{An optical counterpart to the Anomalous X-ray Pulsar
4U~0142+61} 
\author{F. Hulleman\affiliation[1]{Astronomical Institute, Utrecht
University, P.O. Box 80000, 3508~TA Utrecht, The Netherlands},
M. H. van Kerkwijk\affiliationmark[1],
and S. R. Kulkarni\affiliation[2]{Palomar Observatory 105-24,
California Institute of Technology, Pasadena, CA 91125, USA}}
\dates{}{} 
\headertitle{}
\mainauthor{}
\smallskip  

{\it This manuscript has been accepted by Nature. You are free to
refer to this paper in your own paper. However we do place
restrictions on any dissemination in the popular media. The article is
under embargo until it is published (1900 hrs
London time (GMT) on 6 December). For further enquiries, please
contact FH (F.Hulleman@astro.uu.nl) or MvK (M.H.vanKerkwijk@astro.uu.nl).}

\summary{The energy source of the anomalous X-ray pulsars\cite{mere99}
(AXPs) is not well understood, hence their designation as anomalous.
Unlike binary X-ray pulsars, no companions are seen, so the energy
cannot be supplied by accretion of matter from a companion star. The
loss of rotational energy, which powers radio pulsars, is insufficient
to power AXPs. Two models are generally considered: accretion from a large disk left over from the
birth process\cite{marslrh99,chathn00}, or decay of a very strong
magnetic field ($10^{15}\,$G) associated with a
`magnetar.'\cite{thomd96} The lack of counterparts at other
wavelengths has hampered progress, in our understanding of these
objects. Here, we present deep optical observations of the field around
4U~0142+61, which is the brightest AXP in \mbox{X-rays}. The source
has no associated supernova remnant, which, together with its
spin-down time scale of $\simeq10^5\,$yr (ref. \pcite{wilsdf+99}), suggests
that it may be relatively old. We find an object with peculiar optical
colours at the position of the X-ray source, and argue that it is the
optical counterpart. The optical emission is too faint too admit the
presence of a large accretion disk, but may be consistent with
magnetospheric emission from a magnetar.}

\maketitle

The field around 4U~0142+61 was imaged in the V, R and I bands on 31
October 1994 using the Low-Resolution Imaging
Spectrometer\cite{okecc+95} on the 10-m Keck~I telescope on Mauna Kea,
and in R and I on 6 September 1999 on the Keck~II telescope.  In 1994,
the sky was not photometric and the seeing mediocre at 1 arcsec, but
in 1999 the sky was photometric and the seeing 0.6 arcsec. Photometric
calibration was achieved with images obtained under photometric
conditions on 23 July 2000 at the 60-inch telescope on Palomar
Mountain.

Within the error circles of 4U~0142+61 (Figure~\ref{fig:keck}), we find
one faint stellar object with unusual colours (Figure~\ref{fig:ccd}).
To understand its nature, and its relation to 4U~0142+61, we need to
constrain the run of reddening with distance along the line of sight.
We use two open clusters in this region of sky\cite{phelj94} --
NGC~654 (at an angular offset of $17'$, distance of $2.7\pm0.4$\,kpc
and reddening $A_V=2.3\div3.4$) and NGC~663 ($30'$, $2.8\pm0.4$\,kpc,
$2.2\div3.1$) -- to infer that $A_V\simeq2.7$ at $d=2.7\,$kpc.
Furthermore, we use the bluest galaxy in our field (object~G), with
$R=25.6$, $R-I=1.3$.  Because the intrinsic colour\cite{fukusi95} is
unlikely to be bluer than $(R-I)_0=0.3$, the total Galactic reddening
along this line of sight is constrained by $E_{R-I}\simlt1.1$, or
equivalently $A_V\simlt5$.

From Figure~\ref{fig:ccd}, we see that most stars have brightnesses
and colours consistent with those expected for low-mass stars at
distances similar to the open clusters.  Star~A, however, is either
too dim or too blue.  It might be bluer because it is nearer and
therefore less reddened.  But then it would be too dim to be a
main sequence star and too red to be a white dwarf.  The latter can be
seen by comparison to star~D, which is probably a foreground white dwarf
(Figure~\ref{fig:ccd}).

We proceed to estimate the temperature $T$ and radius $R$ of star~A on
the assumption that the object is located beyond the open clusters
(i.e., $d>2.7\,$kpc; $2.7\simlt A_V\simlt5$); below we use $d_5$, the
distance normalized to 5\,kpc.  Bearing in mind the intrinsic
variability in the shape of the reddening curve\cite{cardcm89}, we
find a marginal fit for $T\simeq 6000\,$K and
$R\simeq0.11d_5\,R_\odot$ ($A_V=2.7$), where $R_\odot$ is the Sun's
radius; and increasingly better fits for $T\simeq10^4\,$K and
$R\simeq0.08d_5\,R_\odot$ ($A_V\simeq4$) up to $T\simgt10^5\,$K and
$R\simeq0.022d_5 (T/10^5{\rm\,K})^{-1/2}\,R_\odot$ ($A_V\simeq5.4$).
We note that for $T\simgt10^5$\,K the optical bands fall in the
Rayleigh Jeans tail and the colours do not vary with $T$ any more.
Because temperature and reddening compensate each
other, the above scaling in the Rayleigh-Jeans limit actually holds to
within 10\% for any $T\simgt6000\,$K.

For all temperatures, the radii are too small for star~A to be a
normal star within our Galaxy.  The remaining possibility is that
star~A is a reddened but hot ($T\simeq5\times10^5d_5^2\,$K) white
dwarf.  This is extremely unlikely unless it is actually 4U~0142+61
itself, as we discuss below.

From the above, we conclude that in all likelihood star~A is the
optical counterpart to 4U~0142+61.  Below, we discuss our observations
in the framework of the models that have been proposed for AXPs.  The
optical extinction $A_V$ to 4U~0142+61 is between 2.7 and 5.4\,mag, as
estimated from the X-ray dust scattering halo and 
absorption column density, respectively\cite{whitae+96,preds95} (where
the latter is likely the more reliable\cite{whitae+96}); therefore,
the source must be at a distance exceeding $\sim\!2.7\,$kpc.

We first consider the possibility that AXPs are isolated neutron stars
accreting matter from a disk, presumably composed of supernova
debris\cite{marslrh99,chathn00}.  In this model, the optical emission
arises in the accretion disk, mostly due to reprocessing of the X-ray
irradiation\cite{pernhn00} (Figure~\ref{fig:nufnu}).  The
isotropic X-ray luminosity of 4U~0142+61 is $L_{\rm{}X} =
4\pi{}d^2f_{\rm{}X}^{\rm{}unabs} =
1.8\times10^{36}d_5^2{\rm\,erg\,s^{-1}}$, where
$f_{\rm{}X}^{\rm{}unabs}$ is the flux in the energy range 0.5--10\,keV
after correction for interstellar absorption (see
Table~\ref{tab:compare}).  We can estimate the run of temperature with
disk radius\cite{vrtirg+90} $r$ by $T(r)\simeq5000 (f/0.25)^{2/7}
d_5^{4/7} (r/R_\odot)^{-3/7}{\rm\,K}$.  Here, the factor $f$
parametrizes uncertainties in vertical disk structure and the fraction
of impinging X-ray emission reflected by the disk.

The flux at a given frequency and for given inclination is obtained by
integrating the emission from the accretion disk (assumed to be
optically thick) from an inner radius, $r_{\rm{}in}$, to an outer
radius, $r_{\rm{}out}$; see ref.~\pcite{pernhn00} and
Figure~\ref{fig:nufnu}.  For a simple estimate, we use that the
optical flux will arise predominantly at those radii at which the
emission peaks in the optical band, i.e., where $T\simeq5000\,$K.  For
4U~0142+61, this would be at $r\simeq1 (f/0.25)^{2/3}
d_5^{4/3}\,R_\odot$, which is much larger than the limit of
$R\simeq0.11d_5\,R_\odot$ derived above, and leads to a brightness
much larger than observed (see Figure~\ref{fig:nufnu}).  Thus, this
model is excluded.

The optical emission from the disk can be reduced by appealing to a
larger $r_{\rm{}in}$ or a smaller $r_{\rm{}out}$. The former,
suggested\cite{pernhn00} in the context of earlier
limits\cite{hullvkvk00}, would result in extremely red emission
(because of the absence of the hotter inner region).  This is
incompatible with the observations.  

A reduced $r_{\rm{}out}$ would follow naturally in a model less often
considered, in which the AXP are compact binaries\cite{meres95}.  We
find that our observations require $r_{\rm{}out}\simeq 0.05
d_5^{10/11}(f/0.25)^{-2/11}\,R_\odot$; see Figure~\ref{fig:nufnu}.
This is very small, but not unprecedented: the X-ray binary with the
tightest known orbit\cite{stelwp87}, 4U~1820$-$30 (an 11 min orbital
period) has a similarly small radius and similarly high ratio of X-ray
to optical flux; see Table~\ref{tab:compare}.  However, the X-ray
spectrum of 4U~0142+61 is rather different from that of 4U~1820$-$30
and other accreting sources.  Furthermore, such compact binaries are
not expected to be associated preferentially with supernova remnants.
Thus, we consider this model unlikely.

As mentioned, the optical data are consistent with a hot white dwarf.
This would be expected in another model rarely considered, in which
AXPs are massive and magnetized ($10^8\,$G) white dwarfs rotating on
the verge of breakup, presumably formed in the merger of two ordinary,
approximately $0.6\,M_\odot$ white dwarfs\cite{pacz90}.  In this model, the
ultimate source of energy, as in radio pulsars, is rotation.  For
4U~0142+61, the inferred rotational energy loss, $-\dot{E}_{\rm{}rot}=
4\pi^2 I\dot{P}/P^3 \simeq 10^{37}{\rm\,erg\,s^{-1}}$, is clearly
sufficient to account for the X-ray luminosity; here,
$I=kMR^2\simeq10^{50}{\rm\,g\,cm^2}$ is the moment of inertia
appropriate for a hot white dwarf with mass $M\simeq1.3\,M_\odot$,
radius $R\simeq0.007\,R_\odot$ and gyration constant
$k\simeq0.14$. $P$ is the spin period and $\dot{P}$ its derivative.
If some fraction $\beta$ of $\dot{E}_{\rm{}rot}$ goes into heating, as
might result if the white dwarf rotated differentially, we expect a
surface luminosity $L=4\pi{}R^2\sigma{}T^4= -\beta\dot{E}_{\rm{}rot}$,
and thus a surface temperature $T\simeq4\times10^5 (\beta/0.5)^{1/4}
(k/0.14)^{1/4} (M/1.3\,M_\odot)^{1/4}\,$K (we note that $T$ does not
depend on $R$).  This is sufficient to produce the optical emission
for a source at $d_5=0.6(R/0.007\,R_\odot)$ (see
Figure~\ref{fig:nufnu}).  We note that a possible descendant may
already have been identified\cite{ferrvw+97}: the $1.3\,M_{\odot}$
white dwarf RE~J0317$-$853 with $P=725\,$s, $B\simeq5\times10^8\,$G,
and a cooling age of a few $10^8\,$yr.  In this model,
however, what mechanism may cause the X-ray spectrum is unclear, the thermal
emission being far too soft (see Figure~\ref{fig:nufnu}).
Furthermore, the association of other AXPs with supernova remnants
again seems puzzling. 

This leaves us with the magnetar model\cite{thomd96}, in which both
the power law component of the X-ray emission and the optical emission
would be of magnetospheric origin.  Unfortunately, there are no
detailed models.  Radio pulsars with X-ray and optical detections have
rather smaller $f_{\rm{}X}/f_{\rm{}opt}$ (see
Table~\ref{tab:compare}), but their X-ray spectra are unlike those of
AXPs.  From Figure~\ref{fig:nufnu}, it appears that the simplest
spectral energy distribution would be obtained if the extinction were
near the high end ($A_V\simgt5$), so that the emission could peak at a
few $10^{16}\,$Hz ($\sim\!100\,$eV) and be like a power law
$f_\nu\propto\nu^\alpha$ with index $\alpha\simeq2$ (or even 2.5) in
the optical.  This could arise if the magnetospheric emission is
self-absorbed at optical frequencies.  A possible problem for the
magnetar model is the lack of X-ray emission from the one radio pulsar
with similar $P$ and $\dot{P}$ (ref.~\pcite{pivokc00}). However, it
may well be that, $P$ and $\dot P$ are not reliable measures of the
strength of magnetic fields for AXPs and that other properties
determine whether a source is an AXP.

In spite of the current uncertainties, the optical identification
offer us new insights into these enigmatic objects and motivates new
observations and searches for optical pulsations. 

\clearpage

\bibliographystyle{nature}
\bibliography{u0142}

\section*{Acknowledgements}
We thank D. Kaplan for undertaking the observations at the 60-inch
telescope, and acknowledge helpful discussions with B. Paul and
F. Verbunt.  Model atmospheres for hot white dwarfs were kindly
provided by J. Heise.  The observations were obtained at the
W.~M.~Keck Observatory on Mauna Kea, Hawaii, which is operated by the
California Association for Research in Astronomy.  This research made
use of the SIMBAD data base.  MHvK is supported by the Royal
Netherlands Academy of Science (KNAW).  SRK's research is supported,
in part, by grants from the National Science Foundation and the
National Aeronautics and Space Administration.

Correspondence and requests for materials should be addressed to
F.H. (email:F.Hulleman@astro.uu.nl) 
\clearpage

\begin{table}[t]
\smallskip
\caption[]{Comparison with other high energy objects\label{tab:compare}}
\smallskip

\smallskip\noindent Type of object: AXP: anomalous X-ray pulsar; PSR: radio pulsar; XRB: X-ray
binary; INS: isolated neutron star.\\ 
Although RX~J0720.4$-$3125 has a similar pulse period and X-ray to
optical flux ratio, its X-ray spectrum is very different from those of
the AXPs.\\
The observational data have been collected from the literature.  A
full set of references can be found on
http://www.astro.uu.nl/$\scriptsize\sim$hulleman/u0142.  The unabsorbed X-ray fluxes are for the energy
range of 0.5--10\,keV.  They have been inferred from the best-fit
spectral model with the help of the PIMMS software package. For 4U~1626$-$67,
constant pulsed flux is assumed in the estimate of $(R-I)_0$, 
and for 4U~1820$-$30, we assumed $V-R=0$ to estimate $R$.  For the
conversion to fluxes, we used the Bessell zero points\cite{bess92},
where $V=0$ corresponds to 3600\,Jy, $R=0$ to 3060\,Jy, and $I=0$ to
2420\,Jy.  For reddening corrections, we used $(A_{V_{\rm
L}},A_R,A_I)=(1.015,0.819,0.594)A_V$, where $A_{V_{\rm L}}$, $A_R$,
and $A_I$ are reddenings in the Landolt V, R, and I filters, and $A_V$
is the reddening in the Johnson V band; see Appendix~C in
ref.~\pcite{schlfd98} for a discussion.\\
For reference, the final column lists estimated distances; the colons
indicate the uncertainty, zero colons indicating a factor up to 1.5
uncertainty, one colon a factor two, two colons a factor three.

\end{table}

\clearpage
\begin{table}[t]

\def\phs{\phantom{-}}
\def\phn{\phantom{0}}
\rotate[l]{
\begin{tabular}{@{}ll@{~~}l@{~~}l@{~~~~}l@{~~}l@{~~}l@{~~}r@{~~}r@{~~}rr}	
\hline
Object& Type& 
\multicolumn{1}{c}{\lower5pt\hbox{$\stackrel{\displaystyle P_{\rm
spin}}{(\rm s)}$}}&
$A_R$&
\multicolumn{1}{c}{$\!\!R$}&
\multicolumn{1}{c}{$\!\!\!\!(V\!\!-\!R)_0\!\!$}&
\multicolumn{1}{c}{$\!\!\!(R\!-\!I)_0\!\!$}& 
\multicolumn{1}{c}{\lower5pt\hbox{$\stackrel{\displaystyle f_{\rm X}^{\rm
unabs}}{(\rm erg\,s^{-1}\,cm^{-2})}$}}& 
\multicolumn{1}{c}{$\displaystyle\frac{f_{\rm X}^{\rm unabs}}
{\nu_{R}f^{\rm unabs}_{\nu,R}}$}&
\multicolumn{1}{c}{\lower5pt\hbox{$\stackrel{\displaystyle d}{(\rm
kpc)}$}}
&ref.\\
\hline
4U 0142+61&  AXP&$ 8.69$&$ 4.4$&$ 24.98$&$ -0.33$&$\phs 0.01$&$
6.0\times10^{-10}$& $7.1\times10^3$& $>\!2.7$& \pcite{whitae+96}\\
          &     &       &$ 2.1$&&$\phs0.22$&$\phs0.64$&&
  $5.9\times10^4$\\ 
1E 2259+58.6&AXP&$ 6.97$&$ 3.8$&\llap{$>$}$25.7$&  &   &$
1.8\times10^{-10}$& \llap{$\geq\,$}$7.2\times 10^3$&$\simgt\!5$&\pcite{hullvkvk00},\pcite{rhop97}\\
Crab      &  PSR&$ 0.033$&$ 1.3$&$ 16.21$&$ -0.06$&$\phs0.35$&$
4.0\times10^{-9\phn}$& $2.5\times10^2$&2&\pcite{middpb87},\pcite{beckt97}\\
Vela      &  PSR&$ 0.089$&$ 0.3$&$ 23.93$&$ -0.35$&       &$
7.5\times10^{-12}$& $1.4\times10^3$&0.4\rlap{:}&\pcite{beckt97},\pcite{oegefz93},\pcite{nasumcb97}\\
PSR B0656+14&PSR&$ 0.38$&$ 0.1$&$
24.52$&$\phs0.36$&$\phs0.69$&$1.5\times10^{-12}$& $6.1\times10^2$&0.5\rlap{:}&\pcite{greicf+96},\pcite{kurtsz+98}\\
Geminga   &  PSR&$ 0.24$&$ 0.1$&$ 25.4 $&$ -0.20$&        &$
3.9\times10^{-12}$& $3.6\times10^3$&0.2\rlap{:}&\pcite{beckt97},\pcite{kurtkf+00}\\
4U 1626$-$67&XRB&$ 7.66$&$ 0.2$&$ 18.68$&$ -0.06$&$\phs0.82\rlap{:}$&$
2\!\div\!8\times10^{-10}$& $3\!\div\!14\times10^2$&8\rlap{::}&\phs\pcite{chak98},\pcite{chakbg+97},\pcite{orlaff+98}\\
4U 1820$-$30&XRB& &$ 0.7$&$ 18.87$\rlap{:}&      &
&$\llap{8}\!\div\!16\times10^{-9\phn}$& $1\!\div\!2\times10^4$&8&\pcite{vrtihh+86},\pcite{sosik95}\\
RX J0720.4$-$312\rlap{5}& INS&$ 8.39$&$0.1$&$ 26.9$&     &         &$ 2.5\times10^{-12}$&
  $9.1\times10^3$&$<\!0.4$&\pcite{kulkvk98}\\
\hline
\end{tabular}}

\end{table}

\begin{figure}[p]
\caption[]{Keck images around the X-ray position of 4U~0142+61.  The
large panel shows an overview of the R-band image.  Our candidate, A,
as well as two other blue objects, D and G, are indicated with labels
just to the right of their images.  The small bottom panels are
close-ups of the I-band images of A and G, which show that G is
extended and probably a galaxy, while A is not.  The panels on the right
show the V, R, and I images around star A. The error circle indicates
the X-ray position derived from analysis of Einstein HRI observations
($\alpha_{\rm J2000} = 01^{\rm h}46^{\rm m}22\rlap{.}^{\rm s}03$,
$\delta_{\rm J2000} = +61^\circ 45' 04\rlap{.}''3$, 95\% confidence
level radius of $3.9\,$arcsec; ref.~\pcite{whitmg+87}).  The Einstein
positions have proven to be highly reliable, but for comparison we
display in the I-band panel also the error circles derived from
observations with the EXOSAT LEIT (end figures $22\rlap{.}^{\rm s}40$
and $10\rlap{.}''2$; $4.8\,$arcsec radius; ref.~\pcite{whitmg+87}) and
ROSAT PSPC ($21\rlap{.}^{\rm s}93$, $44' 57\rlap{.}''7$; $6\,$arcsec
radius). The ROSAT position of 4U~0142+61 was derived from archival
data, using a boresight correction inferred from two other X-ray
sources in the field, for which the optical counterparts are
known\cite{motcbb+91,vdbev00} (LS~I~+61$^\circ$ 235 and LP~80-77; our
technique is described in ref.~\pcite{hullvkvk00}).  The astrometry
was done as in ref.~\pcite{hullvkvk00}, by measuring 107 stars from
the USNO-A2.0 catalogue\cite{monebc+00} on a short R-band exposure
(root-mean-square residuals of $0.26\,$arcsec in each coordinate,
consistent with the USNO-A2.0 measurement errors). From the I-band
image, we infer a position of star~A of $\alpha_{\rm J2000}=01^{\rm
h}46^{\rm m}22\rlap{.}^{\rm s}41$, $\delta_{\rm
J2000}=+61^\circ45'03\rlap{.}''2$. We estimate that this position is
on the International Celestial Reference Frame to about 0.2\,arcsec.
The other images give consistent positions.  Our 2$\sigma$ upper limit
to the proper motion is $0.03{\rm\,arcsec\,yr^{-1}}$.
\label{fig:keck}}
\end{figure}

\begin{figure}[p]
\caption[]{Colour-magnitude and colour-colour diagrams for stars near
the position of 4U~0142+61.  For comparison, the magnitudes and
colours expected for stars\cite{baracah98} at the distance, reddening,
and age of the nearby open cluster NGC~654 ($2.7\pm0.4\,$kpc,
$2.7\,$mag, $50\,$Myr; ref.~\pcite{phelj94}; solid line) are shown, as
well as a cooling track for a $0.6\,M_\odot$ white
dwarf\cite{bergwb95} at the same reddening and distance (dashed line).
The arrows indicate the effect of increasing the reddening by
$\Delta{}A_{\rm{}V}=1$.

All optical images were corrected for sensitivity variations using
dome flats, and a small correction for non-linearity of the detector
was applied to the 1999 observations.  For the 1999 I-band
observations, a `fringe' frame was constructed by taking the
second-lowest value for each pixel from seven dithered images, which
was subtracted from the individual images to remove interference
patterns.

Instrumental magnitudes were measured using the DAOPHOT~II
package\cite{stet87} on the 1994 V band and the 1999 R and I-band
images, using only those regions not affected by scattered light from
bright, overexposed stars.  The 1994 V-band images were taken at a
different pointing and rotation angle, which caused different parts on
the sky to be affected by bleed trails, and so on. Therefore, we obtained
V-band measurements only for stars in the central part displayed in
Figure~\ref{fig:keck}.  The photometric calibration was done relative
to the 60-inch data, which were calibrated in turn using exposures of
the standard fields PG~1657$-$042 and NGC~7790\cite{stet00}.  The R
and I calibration was verified using Keck images of the standard field
PG~0231+051\cite{stet00}. We estimate the uncertainties in the zero
points of the magnitudes and colours to be about 0.03\,mag.  For
star~A, we find $R=24.99\pm0.07$, $V-R=0.63\pm0.11$,
$R-I=1.15\pm0.09$.  By comparison with the 1994 images, we find that
its brightness was constant to within 0.2\,mag (2-$\sigma$) in R.
\label{fig:ccd}}
\end{figure}

\begin{figure}[p]
\caption[]{Energy distribution for 4U~0142+61.  At low frequencies
($10^{14}$--$10^{15}$\,Hz), the points marked \texttt{V}, \texttt{R},
and \texttt{I} indicate the observed V, R, and I-band fluxes.  The
vertical error bars reflect the uncertainties, while the horizontal
ones indicate the filter bandwidths.  The set of points above the
measurements indicate dereddened fluxes for $A_V=5.4$, as inferred
from the X-ray column density\cite{whitae+96,preds95}.  The errors
include a 3\% uncertainty in the reddening correction\cite{cardcm89}.

At high frequencies ($10^{17}$--$10^{18}$\,Hz), the crosses show the
incident X-ray spectrum as inferred from \texttt{ASCA}
measurements\cite{whitae+96}.  The diamonds show the spectrum after
correction for interstellar absorption, and the two thick dashed
curves show the two components used in the fit\cite{whitae+96}: a
power law of the form $F_\nu=103(h\nu/1{\rm\,keV})^{-2.67}\,\mu$Jy and
a black body with $T=4.4\times10^6\,$K and $R=12d_5\,$km.  The latter
component is extrapolated to lower frequencies to show it cannot
reproduce the optical fluxes.

Drawn in thin solid lines are models for the optical emission.  The
curve marked \texttt{disk} is for an accretion disk that has inner
radius equal to the corotation radius
$R_{\rm{}co}=(GMP^2/4\pi^2)^{1/3}=0.010\,R_\odot$ (for
$M=1.4\,M_\odot$, $P=8.7\,$s), outer radius of $10^{14}\,$cm, and
inclination of 60 degrees.  Both irradiation and viscous heating are
taken into account\cite{vrtirg+90}; the former dominates in the
optical and the latter causes the bump in the ultra-violet.  The
calculation is for $d=5\,$kpc, but since the disk luminosity scales
almost linearly with the X-ray luminosity, the result is not sensitive
to distance.  The optical fluxes expected in the accretion model are
greatly in excess of those observed.  Only for a truncated disk with a
small outer disk radius,
$r_{\rm{}out}\simlt0.1d_5^{10/11}(f/0.25)^{-2/11}\,R_\odot$, can the
observed fluxes be reproduced, as is shown by the dashed curve.

The thin solid curves marked \texttt{BB}, \texttt{H}, and \texttt{He}
are black body, pure hydrogen and pure helium white dwarf model
atmospheres for $T=4\times10^5\,$K.  The normalization to the optical
data implies $R=0.011d_5$, $0.017d_5$ and $0.015d_5\,R_\odot$,
respectively.  None of the spectra can reproduce the X-ray emission.
\label{fig:nufnu}} 
\end{figure}

\begin{figure}
\centerline{\psfig{figure=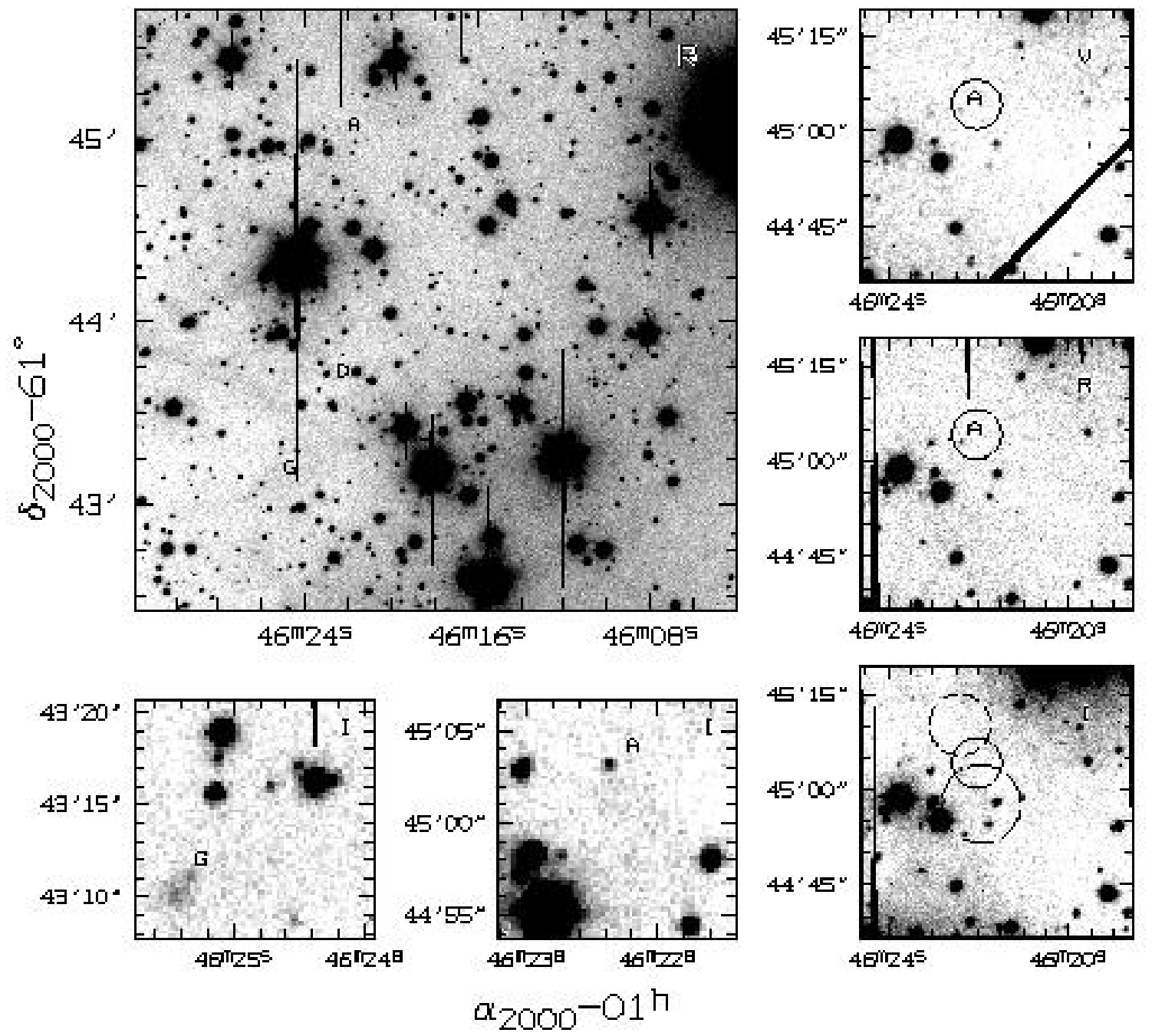,width=\hsize}}
\end{figure}
\begin{figure}
\centerline{\psfig{figure=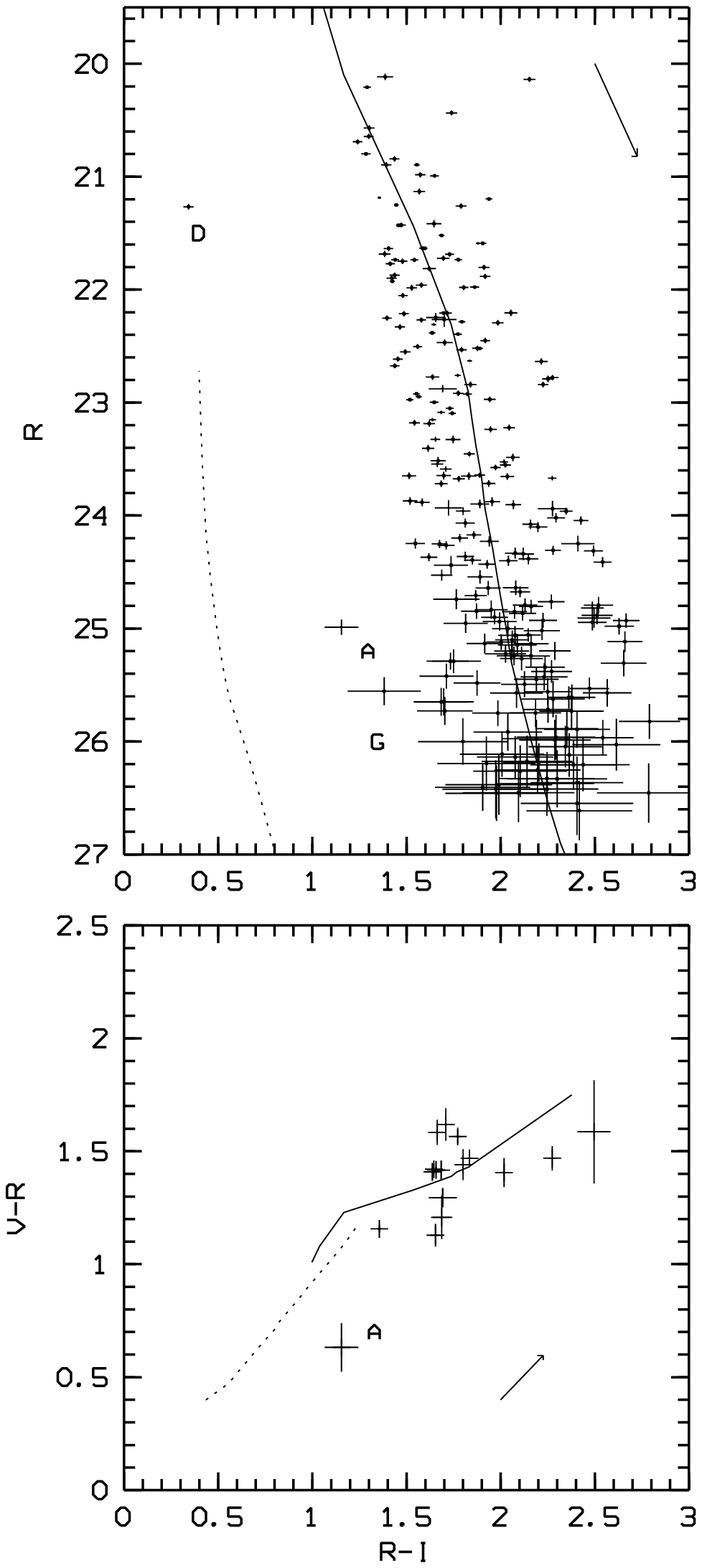,width=0.8\hsize,clip=y}}
\end{figure}

\begin{figure}
\centerline{\psfig{figure=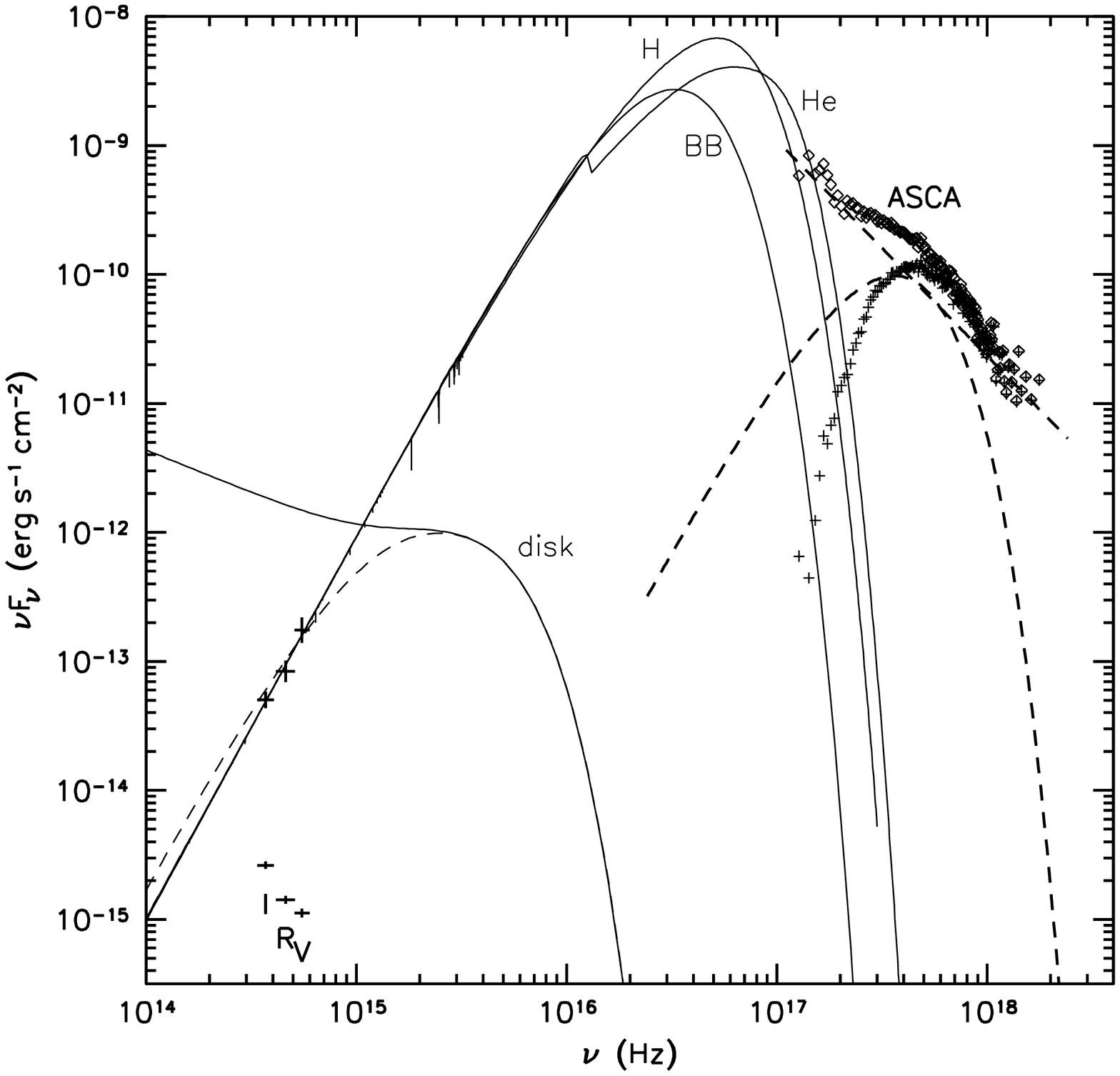,width=\hsize}}
\end{figure}
\end{document}